\documentstyle [12pt]{article}

\newcommand{\be}{\begin{eqnarray}}
\newcommand{\ee}{\end{eqnarray}}

\begin{document}

\centerline{\bf Shock-Wave Cosmology Inside a Black Hole}

\vspace{.3cm}

\centerline{October 17, 2002}
%mistake p32 line 1, ((barr))

$$\begin{array}{ccc}
Joel\ Smoller\footnotemark[1]  &  Blake\ Temple\footnotemark[2] \nonumber \end{array}$$ 
\footnotetext[1]{Department of Mathematics, University of Michigan, Ann Arbor, MI 48109; Supported in
part by NSF Applied Mathematics Grant Number DMS-010-3998, and by the Institute of Theoretical Dynamics,
UC-Davis.}

\footnotetext[2]{Department of Mathematics, University of California, Davis, Davis CA
95616; Supported in part by NSF Applied Mathematics Grant Number DMS-010-2493, and by the Institute of
Theoretical Dynamics, UC-Davis.}

\newtheorem{Theorem}{Theorem}
\newtheorem{Lemma}{Lemma}
\newtheorem{Proposition}{Proposition}
\newtheorem{Corollary}{Corollary}
\small

\begin{abstract}

We construct a class of global exact solutions of the Einstein equations that extend the Oppeheimer-Snyder (OS) model
to the case of non-zero pressure, {\em inside the Black Hole}, by incorporating a shock wave  at the
leading edge of the expansion of the galaxies, arbitrarily far beyond the Hubble length in the Friedmann-Robertson-Walker (FRW) spacetime.  Here the expanding
FRW universe emerges behind a subluminous blast wave that explodes outward from the FRW center at the instant of the Big Bang.  The total mass behind the
shock decreases as the shock wave expands, and the entropy condition implies that the shock wave must weaken to the point where it settles down to an OS
interface, (bounding a {\em finite} total mass), that eventually emerges from the White Hole event horizon of an ambient Schwarzschild spacetime.  The
entropy condition breaks the time symmetry of the Einstein equations, selecting the explosion over the implosion.  These
shock wave solutions indicate a new cosmological model in which the Big Bang arises from a localized explosion occurring inside the Black Hole of an
asymptotically flat Schwarzschild spacetime.

\end{abstract}

\section{Introduction}
\setcounter{equation}{0}

We describe a new cosmological model based on matching a critically expanding FRW metric to a metric which we call the Tolman-Oppenheimer-Volkoff (TOV)
metric {\em inside the Black Hole}, across a shock wave that lies beyond one Hubble length from the center of the FRW spacetime.   This implies that the
spacetime beyond the shock wave must lie {\em inside a Black Hole}, thus extending the shock matching limit of one Hubble length that we identified in
\cite{smolte9}.

In the exact solutions constructed in this paper, the
expanding FRW universe emerges behind a subluminous blast wave that explodes outward from the origin
$\bar{r}=0$ at the instant of the Big Bang $t=0,$
 at a distance beyond one Hubble length\footnotemark[3].\footnotetext[3]{We let $(t,r)$ denote standard FRW coordinates, so that $\bar{r}=rR(t)$ measures
arclength distance at each fixed value of the FRW time $t,$ where $R$ denotes the cosmological scale factor. Barred coordinates also refer to TOV
standard coordinates, in which case  $\bar{r}=rR(t)$ also holds as a consequence of shock matching.}  The shock wave then
continues to weaken as it expands outward until the Hubble length eventually catches up to the shock, and this marks the event horizon of a Black Hole in the
TOV metric beyond the shock.  From this time onward, the shock wave is approximated by a zero pressure ($k=0$) Oppenheimer-Snyder (OS) interface, and thus the
OS solution gives the large time asymptotics of these solutions.  Surprisingly, the equation of state 
$p=\frac{1}{3}\rho$ of early Big Bang physics, is distinguished by the differential equations, and only for this equation of state does the shock wave
emerge from the Big Bang at a finite nonzero speed, the speed of light.  This is surprising because the equation of state $p=\frac{1}{3}\rho$
played no special role in shock matching outside the Black Hole, \cite{smolte2}.  We find it interesting
that such a shock wave emerging from the Big Bang beyond the Hubble length, would thermalize the radiation in a region well beyond the light cone of an
observer positioned at the FRW center, even though the model does not invoke inflation.   Details will appear in our forthcoming paper
\cite{smoltenew}; we wish here to summarize this work, and describe its physical interpretation.

\section{Statement of the Problem}
\setcounter{equation}{0}

If there is a shock wave at the leading edge of the
expansion of the galaxies, then we can ask what is the critical radius
$\bar{r}_{crit}$ at each fixed time $t$ in a $k=0$ FRW metric such that the total mass inside a shock wave positioned beyond that radius puts the universe
inside a Black Hole? (There must be such a critical radius because the total mass
$M(\bar{r},t)$ inside radius
$\bar{r}$ in the FRW metric at fixed time $t$ increases like $\bar{r}^3,$ and so at each fixed time $t$ we must have 
$\bar{r}>2M(\bar{r},t)$ for small enough
$\bar{r},$ while the reverse inequality holds for large $\bar{r}.$ We let 
$\bar{r}=\bar{r}_{crit}$ denote the smallest radius at which
$\bar{r}_{crit}=2M(\bar{r}_{crit},t).$)  We show that when $k=0,$ $\bar{r}_{crit}$ equals
the Hubble length. Thus we cannot match a critically
expanding FRW metric to a classical TOV metric {\em beyond one Hubble length} without continuing the TOV solution into a Black Hole---and we showed in
\cite{smolte3} that the standard TOV metric cannot be continued into a Black Hole.  Thus to do shock matching with a $k=0$ FRW metric beyond one Hubble
length, we introduce the TOV metric {\em inside the Black Hole}. 

\section{The TOV Metric Inside the Black Hole}
\setcounter{equation}{0}

When the metric anzatz is taken to be of the TOV form

\be
ds^2=-B(\bar{r})d\bar{t}^2+A^{-1}(\bar{r})d\bar{r}^2+\bar{r}^2d\Omega^2,
\label{4.1}
\ee
and the stress tensor $T$ is taken to be that of a perfect fluid co-moving with the metric, the Einstein equations $G=\kappa T,$ inside the Black Hole, take
the form
\begin{eqnarray}
\bar{p}'&=&\frac{\bar{p}+\bar{\rho}}{2}\frac{N'}{N-1},\label{4.16}\\ N'&=&-\left\{\frac{N}{\bar{r}}+\kappa
\bar{p}\bar{r}\right\},\label{4.17}\\
\frac{B'}{B}&=&-\frac{1}{N-1}\left\{\frac{N}{\bar{r}}+\kappa\bar{\rho}\right\}.\label{4.18}
\end{eqnarray}
We let $\bar{\rho},$ $\bar{p}$ denote the density and pressure, and $\bar{r}$ is taken to be the timelike variable because we assume 

\be
A(\bar{r})=1-\frac{2M(\bar{r})}{\bar{r}}\equiv 1-N(\bar{r})<0.
\label{4.2}
\ee
Here, $M(\bar{r})$ has the interpretation as the total mass inside the ball of radius $\bar{r}$ when $\bar{r}>2M,$ but $M$ does not have the same
interpretation inside the Black Hole\footnotemark[4] because $\bar{r}<2M.$   
\footnotetext[4]{System (\ref{4.16})-(\ref{4.18}) for $A<0$ differs substantially from the TOV equations for $A>0$ because, for example, the energy
density
$T^{00}$ is equated with the timelike component $G^{rr}$ when $A<0,$ but with $G^{tt}$ when $A>0.$  In particular, this implies that
$M'=-4\pi\bar{p}\bar{r}^2$ when
$A<0,$ versus
$M'=4\pi\bar{\rho}\bar{r}^2$ when $A>0,$ the latter being the equation that gives the mass function its physical interpretation outside the Black Hole.}
The system (\ref{4.16})-(\ref{4.18}) defines the simplest class of gravitational
metrics that contain matter, and evolve {\it inside the Black Hole}.

\section{Shock Matching Inside the Black Hole}
\setcounter{equation}{0} 

Matching a given
$k=0$ FRW metric to a TOV metric {\it inside the Black Hole} across a shock interface, leads to the system of ODE's

\be
\label{2by2first}
\frac{du}{dN}=-\left\{\frac{(1+u)}{2(1+3u)N}\right\}\left\{\frac{(3u-1)(\sigma-u)N+6u(1+u)}{(\sigma-u)N+(1+u)}\right\},\label{ueqn}
\ee
\be
\frac{d\bar{r}}{dN}&=&-\frac{1}{1+3u}\frac{\bar{r}}{N},\label{Neqn}
\ee
with conservation constraint
\be
\label{cc-3}
v=\frac{-\sigma\left(1+u\right)+(\sigma-u)N}{(1+u)+(\sigma-u)N}.
\ee
Here
\be
u=\frac{\bar{p}}{\rho},\ v=\frac{\bar{\rho}}{\rho},\ \sigma=\frac{p}{\rho}, \label{uvdef}
\ee
$\rho$ and $p$ denote the (known) FRW density and pressure, $\sigma=p/\rho,$ and all variables are evaluated at the shock.  Solutions of
(\ref{ueqn})-(\ref{cc-3}) determine the (unknown) TOV metrics that match the given FRW metric Lipschitz continuously across a shock interface, such that
conservation of energy and momemtum hold across the shock, and such that there are no delta function sources at the shock, \cite{isra,smolte3}. 

For such solutions, the speed of the shock interface relative to the fluid comoving on the FRW side of the shock, is given by
\be
s=R\dot{r}=\sqrt{N}\left(\frac{\sigma-u}{1+u}\right). \label{rdot-0}
\label{shock}
\ee
Note that the dependence of (\ref{ueqn})-(\ref{cc-3}) on the FRW metric is only through the variable $\sigma.$   Since solutions of (\ref{ueqn})-(\ref{cc-3})
are formally time-reversible but shock waves are not, system (\ref{ueqn})-(\ref{cc-3}) must be augmented by entropy condition for shocks that breaks the
time symmetry.  Since we are interested in solutions that model the ``Big Bang'' as a localized explosion with an outgoing blast wave
emanating from
$\bar{r}=0$ at time $t=0,$  we impose the entropy conditions,

\be
0<\bar{p}<p,\ \ \ \ \ 0<\bar{\rho}<\rho.
\label{entropy2}
\ee
Condition (\ref{entropy2}) for outgoing shock waves implies that the shock wave is compressive, and is sufficient to rule out
unstable rarefaction shocks in classical gas dynamics.

\section{Exact Shock Wave Solutions Inside the Black Hole}
\setcounter{equation}{0}

In the case when the FRW pressure
is given by the equation of state

\be
p=\sigma\rho,
\label{7.1}
\ee
$\sigma$ assumed constant, $0<\sigma<1,$ the FRW equations have the exact solutions

\be
&\rho=\frac{4}{3\kappa(1+\sigma)^2}\frac{1}{t^2},&\label{frw3-1}\\
 &R=\left(\frac{t}{t_0}\right)^{\frac{2}{3(1+\sigma)}},&\label{frw4-1}
\ee
which assumes an expanding universe, ($\dot{R}>0$), and initial conditions
$R=0$ at $t=0,$ and $R=1,$ at $t=t_0.$  Since
$\sigma$ is constant, equation (\ref{ueqn}) uncouples from (\ref{Neqn}), and thus solutions of system (\ref{ueqn})-(\ref{cc-3}) are
determined by the scalar non-autonomous equation (\ref{ueqn}).  Making the change of variable $S=1/N,$ which transforms 
the ``Big Bang''
$N\rightarrow \infty$ over to rest point at $S\rightarrow0,$ c.f. \cite{smolte9}, we obtain, 

\be
\frac{du}{dS}=\left\{\frac{(1+u)}{2(1+3u)S}\right\}\left\{\frac{(3u-1)(\sigma-u)+6u(1+u)S}{(\sigma-u)+(1+u)S}\right\},\label{-6-1}\label{12}
\ee
We take as entropy condition (\ref{entropy2}), and in addition, we require that the TOV equation of state meet the physical bound
\be
&0<\bar{p}<\bar{\rho}.&\label{state1}
\ee   
Note that the conditions $N>1$ and $0<\bar{p}<p$ restrict the domain of
(\ref{-6-1}) to the region
$0<u<\sigma<1,$ $0<S<1.$   Inequalities (\ref{entropy2}) and (\ref{state1}) are both implied by the single condition
\be
\label{6.5}
S<\left(\frac{1-u}{1+u}\right)\left(\frac{\sigma-u}{\sigma+u}\right).
\ee 
We prove the following theorem:

\begin{Theorem}\label{theorem6}
For every $\sigma,$ $0<\sigma<1,$ there exists a unique solution $u_{\sigma}(S)$ of (\ref{-6-1}), such that (\ref{6.5}) holds on the
solution for all $S,$ $0<S<1,$  and on this solution,
$
0<u_{\sigma}(S)<\bar{u},$ 
$\lim_{S\rightarrow0}u_{\sigma}(S)=\bar{u},
$
where
$
\bar{u}=Min\left\{1/3,\sigma\right\}.$
and
\be
\lim_{S\rightarrow1}\bar{p}=0=\lim_{S\rightarrow1}\bar{\rho}.&\label{Sarrow1}
\ee
\end{Theorem}

\noindent Concerning the the shock speed, we have:

\begin{Theorem}\label{theorem7}
Let $0<\sigma<1.$  Then the shock wave is everywhere subluminous, that is, the shock speed $s_{\sigma}(S)\equiv s(u_{\sigma}(S))<1$ for all $0<S\leq1,$ if and
only if 
$
\sigma\leq 1/3. $
\end{Theorem}
By a careful analysis of the asymptotics of the solution near $S=0,$ we can prove

\begin{Theorem}\label{thm8}
The shock speed at the Big Bang $S=0$ is given by:
\be
\label{shockspeed1}
\lim_{S\rightarrow0}s_{\sigma}(S)=0,\ \ \ \sigma<1/3,
\ee
\be
\label{shockspeed3}
\lim_{S\rightarrow0}s_{\sigma}(S)=\infty,\ \ \ \sigma>1/3,
\ee
\be
\label{shockspeed2}
\lim_{S\rightarrow0}s_{\sigma}(S)=1,\ \ \ \sigma=1/3,
\ee

\end{Theorem} 

\section{Bounds on the Shock Position}
\setcounter{equation}{0}

Let $r_*$ be the FRW radial position of the shock wave at the instant of the Big Bang, (the arclength distance $\bar{r}_*=r_*R(0)=0$ when $R(0)=0$). The
analysis implies that the shock wave will first become visible at the FRW center
$\bar{r}=0$ at the moment $t=t_0,$ ($R(t_0)=1$), when the Hubble length $H_0^{-1}=H^{-1}(t_0)$ satisfies

\be
\frac{1}{H_0}=\frac{1+3\sigma}{2}r_*,
\ee
where $r_*$ is the FRW position of the shock at the instant of the Big Bang.   At this time, we prove that the number of
Hubble lengths $\sqrt{N}_0$ from the FRW center to the shock wave at time $t=t_0$ can be estimated by 
\be
\nonumber
1\leq\frac{2}{1+3\sigma}\leq \sqrt{N}_0\leq \frac{2}{1+3\sigma}e^{\sqrt{3\sigma}\left(\frac{1+3\sigma}{1+\sigma}\right)}.
\ee
Thus, in particular, the shock wave will
still lie beyond the Hubble length
$1/H_0$ at the FRW time $t_0$ when it first becomes visible.  Furthermore, we prove that the time $t_{crit}>t_0$ at which the shock wave will emerge from the
White Hole given that $t_0$ is the first instant at which the shock becomes visible at the FRW center, can be estimated by

\be
\label{finalestimate1}
\frac{2}{1+3\sigma}e^{\frac{1}{4}\sigma}\leq \frac{t_{crit}}{t_0}\leq
\frac{2}{1+3\sigma}e^{\frac{2\sqrt{3\sigma}}{1+\sigma}},
\ee
for $0<\sigma\leq 1/3,$ and by the better estimate
\be
\label{finalestimate2-22}
e^{\frac{\sqrt{6}}{4}}\leq \frac{t_{crit}}{t_0}\leq
e^{\frac{3}{2}},
\ee
in the case $\sigma=1/3.$
For example, (\ref{finalestimate1}), (\ref{finalestimate2-22}) imply that at the OS limit $\sigma=0,$ 
$$\sqrt{N_0}=2,\ \  \frac{t_{crit}}{t_0}=2,$$ 
and in the limit $\sigma=1/3,$
$$1.8\leq\frac{t_{crit}}{t_0}\leq 4.5,\ \ 1<\sqrt{N_0}\leq 4.5,\ \ $$

We conclude in these shock wave cosmological models, that the moment $t_0$ when the shock wave first becomes visible at the
FRW center, it must lie within 4.5 Hubble lengths of the center.  Throughout the expansion up until this time, the expanding
universe must lie entirely within a {\em Black Hole}---the universe will eventually emerge from this {\em Black Hole}, but not until some later time
$t_{crit},$ where $t_{crit}$ does not exceed $4.5t_0.$

\section{Concluding Remarks}
\setcounter{equation}{0}

We have constructed global exact solutions of the Einstein equations in which the expanding FRW
universe extends out to a shock wave that lies arbitrarily far beyond the Hubble length.  The distance to the shock wave at any given value of the Hubble
constant is determined by one free paramenter which can be taken to be the FRW position of the shock wave at the instant of the Big Bang. 

The critical OS solution {\em inside the Black Hole} is
obtained in the limit of zero pressure, and provides the large time asymptotics, but the shock wave solutions differ qualitatively from the OS solution.  For
example, the shock wave models contain matter, and thus do not rely on any portion of the empty space Schwarzschild metric inside the Black Hole for their
construction.  In both models, the interface propagates outward from the FRW center $\bar{r}=0$ at the instant of the Big Bang, but in the shock wave model
the mass function
$M(\bar{r},t)$ is infinite at the instant of the Big Bang, and immediately becomes a finite decreasing function of FRW time, for all future times
$t>0.$  And although the OS solution is time reversible, the directional orientation of the shock interface relative to the various observers is determined by
an entropy condition, \cite{lax,glim,smol}.  The entropy condition selects the explosion over the implosion, and the condition that the entropy condition be
satisfied globally, determines a {\em unique} solution.
%\footnotemark[5]
%\footnotetext[5]{ The time orientation of a
%solution must be selected based on an extra condition, such as an entropy condition for shocks, because the Einstein
%equations and the compressible Euler equations, taken by themselves, are both time reversible, \cite{lax,glim,smol}.  
%Thus the
%entropy condition for the shock is what determines the time orientation for the global dynamics of the solutions we
%construct: in our examples, the FRW metric expanding outward behind a shock wave emanating from a White Hole is entropy satisfying, while its time
%reversal, the FRW metric contracting into a Black hole, is entropy violating.} 
Since the TOV radial coordinate
$\bar{r}$ is timelike inside the Black Hole, we can also say that the density
$\bar{\rho}(\bar{r})$ and mass
$M(\bar{r})$ are both constant {\it at each fixed time} in the TOV spacetime beyond the shock wave. 

The shock interface that marks the boundary of the FRW expansion continues out through the White Hole event horizon of an
ambient Schwarzschild metric at the instant when the wave is exactly one Hubble length from the FRW center
$\bar{r}=0,$ and then continues on out to infinity along a geodesic of the Schwarzschild metric outside the Black Hole. 
These solutions thus indicate a scenario for the Big Bang in which the expanding universe emerges from an explosion
emanating from a White Hole singularity inside the event horizon of an asymptotically flat Schwarzschild spacetime of
finite mass.  The model does not require the physically implausible assumption that the uniformly expanding portion of the
universe is of infinite mass and extent at every fixed time, and it has the nice feature that it embeds the Big Bang
singularity of cosmology within the event horizon of a larger spacetime, the Schwarzschild spacetime.  Moreover, the model
also allows for the uniform expansion of arbitrarily large densities within an arbitrarily large mass extended over an arbitrary number of Hubble lengths
early on in the Big Bang, a prerequisite for the standard physics of the Big Bang at early times.

We conclude that these shock wave solutions give the global dynamics of strong
gravitational fields in an exact solution, the dynamics is qualitatively different from the dynamics of solutions when the
pressure
$p\equiv0,$ the solution suggests a Big Bang cosmological model in which the expanding universe is bounded throughout its
expansion, and the equation of
state most relevant at the Big Bang,
$p=\frac{1}{3}\rho,$ is distinguished by the differential equations.  But these solutions are only rough qualitative
models because the equation of state on the TOV side is determined by the equations, and therefore cannot
be imposed. That is, the TOV density $\bar{\rho}$ and pressure $\bar{p}$ only satisfy the entropy conditions (\ref{entropy2}),
together with the loose
physical bounds
(\ref{state1}); and on the FRW side, the equation of state is taken to be
$p=\sigma\rho,$ $\sigma\equiv const.,$ $0<\sigma<1.$   Nevertheless, these
bounds imply that the equations of state are qualitatively reasonable, and we expect that these solutions will capture the gross dynamics of solutions when
more general equations of state are imposed.  For more general equations of state, other waves, (e.g. rarefaction waves), would need to be present to meet
the conservation constraint, and thereby mediate the transition across the shock wave.   Such transitional waves would be pretty much impossible to model in
an exact solution, but the fact that we can find global solutions that meet our physical bounds, and that are qualitatively the same, for all values of
$\sigma\in (0,1),$ and all initial shock positions, leads us to expect that such a shock wave would be the dominant wave in a large class of problems.

As a final remark, we note that because Einstein's theory by itself does not choose an
orientation for time,  it
follows that if we believe that a Black Hole can exist in the forward time collapse of a mass through an event horizon as
$t\rightarrow\infty,$ (the time $t$ as observed in the far field), then we must also allow for the possibility of the time
reversal of this process, a White Hole explosion of matter out through an event horizon coming from
$t\rightarrow -\infty.$   That is, if we are willing to accept Black Holes in the forward time dynamics of astrophysical objects whose
collapse appears so great as to form an event horizon in the future, then by symmetry, we may well also be forced to
accept White Holes in the backward time dynamics of astrophysical objects which, like the expanding universe, appear to have expansions so great as to 
have emerged from an event horizon in the past.  Given this, it is natural to wonder if there
is a connection between the mass that disappears into Black Hole singularities, and the mass that emerges from White Hole singularities.


\begin{thebibliography}{99}

\bibitem{glim}  J. Glimm, {\em Solutions in the large for nonlinear hyperbolic
systems of equations}, Comm. Pure Appl. Math., {\bf 18}(1965), pp. 697-715.

\bibitem{isra}  W. Israel, {\it Singular hypersurfaces and thin shells 
in General Relativity\/}, IL Nuovo Cimento, Vol. XLIV B, N. 1, 1-14 (1966).

\bibitem{lax}  P.D. Lax, {\it Hyperbolic systems of conservation 
laws, II\/}, Comm.
Pure Appl. Math., {\bf 10}, pp. 537--566 (1957).

\bibitem{smol} J. Smoller, {\it Shock-Waves and Reaction-Diffusion Equations\/}, Springer
Verlag, 1983.

\bibitem{smolte2} J. Smoller and B. Temple {\it Astrophysical shock-wave solutions of the Einstein
equations}, Phys. Rev. D, {\bf 51}, No. 6, 2733-2743 (1995).

\bibitem{smolte3} J. Smoller and B. Temple {\it Shock-wave solutions of the Einstein equations that
extend the Oppenheimer-Snyder model}, Arch. Rat. Mech. Anal., 138, 239-277 (1997).

\bibitem{smolte9}  J.~Smoller and B.~Temple, {\it Cosmology with a shock-wave}, Comm. Math. Phys., 210, 275-308 (2000).

\bibitem{smoltenew} J. Smoller and B. Temple, {\em Cosmology, black holes, and shock waves beyond the Hubble length}, (preprint). 

\bibitem{wein} S. Weinberg, {\em Gravitation and Cosmology: Principles and Applications of the
General Theory of Relativity}, John Wiley \& Sons, New York, 1972.

\end{thebibliography}
\end{document}